\documentstyle[prl,aps,epsfig,multicol]{revtex}
\begin{document}
\def\be{\begin{equation}}
\def\ee{\end{equation}}
\def\bearr{\begin{eqnarray}}
\def\eearr{\end{eqnarray}}
\def\tc{$T_c~$}
\def\tcl{$T_c^{1*}~$}
\def\c2{ CuO$_2~$}
\def\ruo{ RuO$_2~$}
\def\lsco{LSCO~}
\def\bi{bI-2201~}
\def\tl{Tl-2201~}
\def\hg{Hg-1201~}
\def\sro{$Sr_2 Ru O_4$~}
\def\rc{$RuSr_2Gd Cu_2 O_8$~}
\def\mgb{$MgB_2$~}
\def\pz{$p_z$~}
\def\ppi{$p\pi$~}
\def\sqo{$S(q,\omega)$~}
\def\tperp{$t_{\perp}$~}

\title{Reply to a comment by Peres et al.}

\author{ G. Baskaran$^1$ and S.A. Jafari$^{1,2}$\\
${}^1$Institute of Mathematical Sciences, 
C.I.T. Campus, Chennai 600 113, India \\
${}^2$Department of Physics, Sharif University of Technology,
Tehran 11365-9161, Iran}

\maketitle
\begin{abstract}

\end{abstract}

\begin{multicols}{2}[]

Peres et al. comment\cite{peres} that in our recent work on 
graphite\cite{gbak1} we get spin-1 collective (spin-1 zero sound)
mode only when we use, in addition to `random phase approximation' (RPA), 
another approximation. 

To justify our method and also clarify the issue involved, {\bf 
particularly at low energies}, we discuss our 
spin-1 zero sound in a fermi liquid theory framework\cite{gbak2} and 
emphasize an important aspect of Landau's zero sound collective mode 
analysis. One calculates 
the two particle forward scattering amplitude defined by Landau 
$\Gamma(p_1,p_2;k)$ by focusing on the singularity in the zero sound 
channel. Most importantly, as emphasized by Landau\cite{landau} one considers 
intermediate states in which the number of `quasi-particles' are not changed. 
The RPA susceptibility we have computed in our approximation precisely 
corresponds to a bubble sum of such a process, where the quasi-particle 
number is conserved in the intermediate states. 

To elaborate this point we write the `bare' Hubbard Hamiltonian for the 
ideal graphene sheet in terms of the conduction band electron ($e'$s)
and valence band hole ($h$'s) operators with {\bf positive energies}:
\bearr
H & = & \sum_{k \sigma}  \epsilon_k ( e^{\dagger}_{k\sigma} e^{}_{k\sigma} + 
h^{\dagger}_{k\sigma} h^{}_{k\sigma}) \nonumber \\
&+& \frac{U}{N} \sum_{k k'}
\gamma^{(1)}_{k,k',q} (e^{\dagger}_{k\uparrow} e^{}_{k+q \uparrow}  
h^{\dagger}_{k'\downarrow} h^{}_{k'-q\downarrow} + ..)\nonumber \\
&+& \frac{U}{N} \sum_{k k'} ( \gamma^{(2)}_{k,k',q}e^{\dagger}_{-k\uparrow}
h^{\dagger}_{k+q\uparrow} e^{\dagger}_{-k'\downarrow} 
h^{\dagger}_{k'-q\downarrow} \nonumber \\
&+&
\gamma^{(3)}_{k,k',q}e^{\dagger}_{k\uparrow}e^{}_{k-q\uparrow}
e^{\dagger}_{-k'\downarrow} h^{\dagger}_{k'-q \downarrow + ..}
\eearr
Here $ \epsilon_k \equiv t \sqrt{ 1 + 4 \cos \frac{{\sqrt 3}k_x}{2} 
\cos \frac{k_y}{2} + 4 \cos^2 \frac{k_y}{2} } $ and
$\gamma$'s are complex form factors, $|\gamma| \sim 1$.
Hubbard interaction terms describe various two body processes:
electron-electron, hole-hole, electron-hole scattering, electron
decay, vacuum polarization etc.
The second line of eqn.(1), $e^{\dagger}e^{}h^{\dagger}h^{}$
is the one important for us - it represents scattering between a real 
electron and a real hole. The term $ e^{\dagger}e^{\dagger}
h^{\dagger}h^{\dagger}$ of the third line corresponds to 
spontaneous creation of two electron-hole pairs out of the vacuum and 
the second term $e^{\dagger}e^{}h^{\dagger}e^{\dagger}$ represents
decay of a real electron into an electron and electron-hole pair. 
That is, all terms contained in the third line are either processes 
that renormalize the vacuum or dress the quasi-particles. 
At the end, if a fermi liquid state is formed, one has `stable' low energy 
renormalized {\bf quasi particles that neither feel the vacuum polarization 
processes nor polarize the vacuum}. 
In a renormalization group sense, the $\gamma^{(2)}$, $\gamma^{(3)}$ etc. 
terms of eqn.(1) are expected to renormalize to zero, as one approaches a
fermi liquid fixed point. For our analysis of spin-1 zero sound in graphite, 
assumed to be a fermi liquid, we are essentially left with the $\gamma^{(1)}$ 
term with a renormalized ${\tilde U} > 0$. That is, one low energy electron 
and one hole quasi-particle repeatedly scatter, in a renormalized 
{\bf frozen fermi liquid vacuum}, 
without intervention of any additional quasi-particles 
in the intermediate state, in the way described by Landau. 
This leads to our simple susceptibility expression\cite{gbak1} 
\be
\chi(q,\omega) = 
\frac{\chi_0 (q,\omega)}{1 - {\tilde U} \chi_0(q,\omega)}
\ee
that has poles corresponding to our predicted spin-1 collective mode 
branch, below the particle-hole continuum. 

\begin{figure}[h]
\epsfxsize 6cm
\centerline {\epsfbox{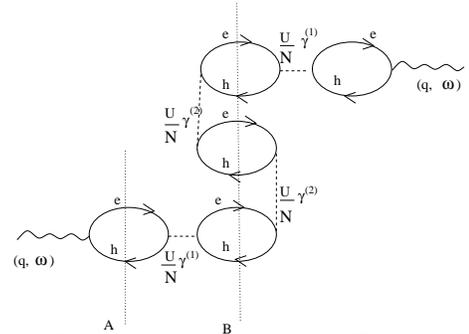}}
\caption{A diagram that is contained in Peres et al.'s bubble summation.
Intermediate state A is a 1-electron and 1-hole state; B is a 3-electron 
and 3-hole state. At low energies, in a Landau theory such diagrams with 
more than 1-electron and 1-hole in the intermediate states do not exist, 
because $U\gamma^{(2)}$ is renormalized to zero/low values.} 
\end{figure}

Peres et al., have, on the other hand, calculated a `2-sub lattice 
susceptibility tensor', which uses the full unrenormalized Hubbard 
interaction terms 
[lines 2,3 and 4 in eqn.(1)] in their bubble summation procedure. 
When converted into a physically more transparent valence and 
conduction bond quasi-particle basis of our fermi liquid state, Peres 
et al.'s susceptibility contains diagrams in which intermediate states 
have higher quasi-particle numbers (an example is shown in figure 1). 
Once we {\bf isolate and remove 
these processes from their bubble summation}, we recover our fermi liquid
result (equation 2) and a full spin-1 collective mode branch emerges 
from the bottom of the particle-hole continuum.

\end{multicols}
\end{document}